\documentclass{Nanotech} 
\usepackage{epsfig}

\title{Modeling the Field Emission Current Fluctuation in Carbon Nanotube Thin Films}

\author{N. Sinha*, D. Roy Mahapatra**, J.T.W. Yeow* and R. Melnik***\\\\
{* Department of Systems Design Engineering, University of Waterloo,Waterloo, ON, Canada}\\
{** Department of Aerospace Engineering, Indian Institute of Science, Bangalore, India}\\ 
{*** Mathematical Modeling and Computational Sciences,Wilfrid Laurier University, 
Waterloo, ON, Canada}\\ 
{jyeow@engmail.uwaterloo.ca}\\}

\begin{document}

\maketitle

\section*{ABSTRACT}

Owing to their distinct properties, carbon nanotubes (CNTs) 
have emerged as promising candidate for field emission 
devices. It has been found experimentally that the results 
related to the field emission performance show variability. 
The design of an efficient field emitting device requires the 
analysis of the variabilities with a systematic and multiphysics 
based modeling approach. In this paper, we develop a model of 
randomly oriented CNTs in a thin film by coupling the field 
emission phenomena, the electron-phonon transport and the 
mechanics of single isolated CNT. A computational scheme is 
developed by which the states of CNTs are updated in time 
incremental manner. The device current is calculated by using 
Fowler-Nordheim equation for field emission to study the 
performance at the device scale.


\section{INTRODUCTION}

\begin{figure}
\begin{center}
\epsfig{file=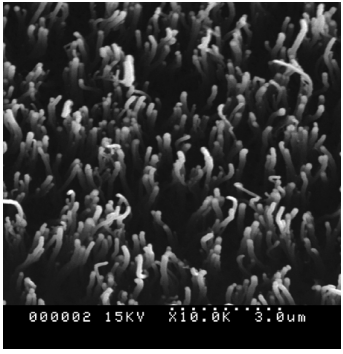, width=5cm, height=5cm}
\end{center}
\label{fig:sem2}
\caption{SEM image showing randomly oriented tips 
of CNTs in a thin film.}
\end{figure}

Field emission from carbon nanotubes (CNTs) was first reported in 
1995~\cite{bib:rinzler},\cite{bib:heer}. 
With advancement in synthesis techniques, application of CNTs in field 
emission devices, such as field emission displays, gas discharge tubes 
and X-ray tube sources has been successfully demonstrated~\cite{bib:bonard},
\cite{bib:saito}. Field emission 
performance of a single isolated CNT is found to be remarkable due to its 
structural integrity, geometry, chemical stability and high thermal 
conductivity. One can use a single CNT to produce an electron beam in a 
single electron beam device. However, in many applications (such as X-ray 
imaging systems), a continuous or patterned film is required to produce 
several independent electron beams. However, the situation in these 
cases becomes complex due to coupling among (i) the ballistic electron-
phonon transport at moderate to high temperature range, (ii) field 
emission from each of the CNT tip and (iii) electrodynamic forces causing 
mechanical srain and deforming CNTs (and thus changing the electron density 
and dynamic conductivity). In such cases, the individual CNTs are 
not always inclined normal to the substrate surface (as shown in 
Fig.~\ref{fig:sem2}, 
where CNT tips are oriented in a random manner). This is the most common 
situation, which can evolve from an initially ordered state of uniformly 
distributed and vertically oriented CNTs. Such evolution process 
must be analyzed accurately from the view point of long-term performance 
of the device. The interest of 
the authors' towards such an analysis 
and design studies stem from the problem of precision biomedical 
X-ray generation.\par 
In this paper, we focus on a diode configuration, where the cathode 
contains a CNT thin film grown on a metallic substrate and 
the anode is a copper plate acting as emission current collector. 
Here, the most important requirement is to have a stable field emission 
current without compromising the lifetime of the device. As the CNTs 
in the film deplete with time, which is due to burning and fragmentation 
that result in a decreasing number of emitting sites, one observes 
fluctuation in the output current. Small spikes in the current have 
also been observed experimentally~\cite {bib:bonard1}, which can 
generally be attributed to the change in the gap between the 
elongated CNT tip and the anode, and also possibly a dynamic contact 
of pulled up tip with the anode under high voltage. As evident from 
the reported studies~\cite {bib:bonard1}, it is 
important to include various coupled phenomena in a numerical model, 
which can then be employed to understand the effects of various 
material parameters and also geometric parameters (e.g. CNT 
geometry as well as thin film patterns) on the collective field 
emission performance of the thin film device. Another aspect of 
interest in this paper is the effect of the angle of orientation 
of the CNT tips on the collective performance. A physics 
based modeling approach has been developed here to analyze the 
device-level performance 
of a CNT thin film. \par

\section{MODEL FORMULATION}


The physics of field emission from metallic surfaces is fairly well understood. The
current density ($J$) due to field emission from a metallic surface is usually
obtained by using the Fowler-Nordheim (FN) equation~\cite {bib:fowler} 
\begin{equation}
J=\frac {B E^{2}}{\Phi} \exp 
\bigg[-\frac {C \Phi ^{3/2}}{E}\bigg] \;,
\label{eq:FN}
\end{equation}
where E is the electric field, $\Phi$ is the work function of the cathode material, 
and B and C are constants. However, in the case of 
a CNT thin film acting as cathode, the surface of the cathode is not smooth 
(like the metal emitters) and consists of hollow tubes in curved shapes and with 
certain spacings. An added complexity is 
the realignment of individual CNTs due to electrodynamic interaction between the 
neighbouring
CNTs during field emission. Analysis of these processes requires the determination 
of the current density by considering the individual geometry of the CNTs, their 
dynamic orientations and the variation in the electric field during 
electronic transport. 

\begin{figure}
\begin{center}
\epsfig{file=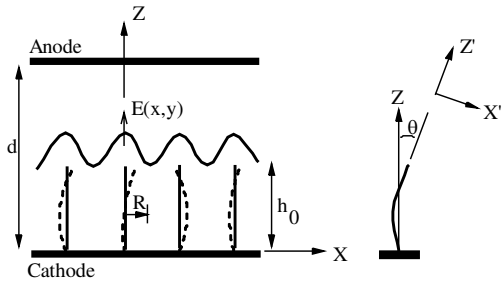, width=7cm}
\end{center}
\label{fig:Exy}
\caption{CNT array configuration.}
\end{figure}


Based on our previously developed model~\cite {bib:sinha1}, which describes 
the degradation 
of CNTs and the CNT geometry and orientation, the rate of degradation of 
CNTs is defined as 
\begin{equation}
v_{\rm burn}=V_{\rm cell} \frac {dn_{1}(t)}{dt} 
\bigg[\frac {s(s-a_{1})(s-a_{2})(s-a_{3})} 
{n^2a_{1}^2+m^2a_{2}^2+nm(a_{1}^2+a_{2}^2-a_{3}^2)} \bigg]^{1/2} \;,
\label{eq:damp2}
\end{equation}
where $V_{\rm cell}$ is the representative volume element, 
$n_{1}$ is the concentration of carbon atoms in the cluster 
form in the cell, $a_{1}$, $a_{2}$, $a_{3}$ are lattice constants, 
$s=\frac {1}{2}(a_{1}+ a_{2}+a_{3}), $$n$ and $m$ are integers 
$(n \geq |m| \geq 0)$. The pair 
$(n,m)$ defines the chirality of the CNT. Therefore, at a given time, 
the length of a CNT can be expressed as
$h(t)=h_0-v_{\rm{burn}}t$, where $h_{0}$ is the initial average 
height of the CNTs and $d$ is the distance between the cathode 
substrate and the anode (see Fig. 2). \par
The effective electric field component for field emission 
calculation in Eq.~(\ref{eq:FN}) is expressed as 
\begin{equation}
E_{z}= -e^{-1} \frac{d{\mathcal V}(z)}{dz} \;,
\label{eq:damp176} 
\end{equation}
where $e$ is the positive electronic charge and 
$\mathcal V$ is the electrostatic potential energy. The total 
electrostatic potential energy can be expressed as 
\begin{equation}
{\mathcal V}(x,z) = -eV_s - e(V_d-V_s)\frac{z}{d} + 
\sum_{j} G(i,j)(\hat{n}_j-n) \;,
\label{eq:E1}
\end{equation}
where $V_s$ is the constant source potential (on the 
substrate side), $V_d$ is the drain potential (on the anode side), 
$G(i,j)$ is the Green's function~\cite{bib:anant} 
with $i$ being the ring position, $\hat{n}_j$
denotes the electron density at node position $j$ on the ring, 
and ($n,m$) denotes the chirality parameter of the CNT. 
The field emission current ($I_{\rm{cell}}$) 
from the anode surface
associated with the elemental volume $V_{\rm{cell}}$ of the film is 
obtained as 
\begin{equation}
I_{\rm{cell}} = A_{\rm{cell}}\sum_{j=1}^{N} J_j \;,
\label{eq:E2}
\end{equation}
where $A_{\rm{cell}}$ is the anode surface area and $N$ is the
number of CNTs in the volume element. The toal current is obtained by summing 
the cell-wise current ($I_{\rm{cell}}$). This formulation takes into account 
the effect of CNT tip orientations, 
and one can perform statistical analysis of 
the device current for randomly distributed and randomly oriented CNTs. 
However, due to the 
deformation of the CNTs due to electrodynamic forces, 
the evolution process requires a much more detailed treatment from the
mechanics point of view.\par
Based on the studies reported in 
published literature, it is reasonable to expect that a major contribution 
is by the Lorentz force due to the flow of electron gas along the 
CNT
and 
the ponderomotive force due to electrons in the oscillatory electric 
field
. The oscillatory electric field could be due to 
hopping of the electrons along the CNT surfaces and the changing 
relative distances between two CNT surfaces.
In addition, the electrostatic force and the van der Waals force are 
also important. The net force components acting on the CNTs parallel 
to the $Z$ and the $X$ 
directions are calculated as~\cite{bib:sinha2} 
\begin{equation}
f_{z}= \int (f_{lz}+f_{vs_{z}})ds+ f_{c_{z}}+ f_{p_{z}} \;,
\label{eq:damp35}
\end{equation}
\begin{equation}
f_{x}= \int (f_{lx}+f_{vs_{x}})ds+ f_{c_{x}}+ f_{p_{x}} \;.
\label{eq:damp36}
\end{equation}
where $f_{l}$, $f_{vs}$, $f_{c}$ and $f_{p}$ are Lorentz, van der Waals, 
coulomb and ponderomotive forces, respectively and $ds$ is the length of 
a small segment of CNTs. 
Next, we employ these force components in the expression of 
work done on the ensemble of CNTs
and formulate an energy conservation law. Due to their large aspect 
ratio, the CNTs have been idealized as one-dimensional elastic 
members (as in Euler-Bernoulli beam). By introducing the 
strain energy density, the kinetic energy density and
the work density, and applying the Hamilton principle, we obtain 
the governing equations in ($u_{x'}$, $u_{z'}$) for each CNT, which can be 
expressed as  
\[
E'A_{2} \frac {\partial^{4} u_{x'}^{(m)}}{\partial z'^4}+ \rho A_{0} {\ddot u_{x'}^{(m)}}-\rho A_{2} \frac {\partial^{2} {\ddot u_{x'}^{(m)}}}{\partial z'^2}-\sum_{m} \pi C_{vs} [({r^{(m+1)}})^2
\]
\begin{equation}
-({r^{(m)}})^2]
\frac {1}{ \Delta_{x'}} \frac {\partial \Delta_{x'}}{\partial z'}\cos 
\big(\theta (z')\big)-f_{lx'}-f_{cx'}=0 \;,
\label{eq:damp49}
\end{equation}
\[
-E'A_{0} \frac {\partial^{2} u_{z'0}^{(m)}}{\partial z'^2}- 
\frac {1}{2}E'A_{0} 
\alpha \frac {\partial \Delta T(z')}{\partial z'}+\rho A_{0}  
{\ddot u_{z'0}^{(m)}}-\pi C_{vs} 
\]
\begin{equation}
[({r^{(m+1)}})^2-({r^{(m)}})^2]
\frac {1}{ \Delta_{x'}} \frac {\partial \Delta_{x'}}{\partial z'} 
\sin \big (\theta (z')\big) -f_{lz'}-f_{cz'}=0 \;,
\label{eq:damp50}
\end{equation}
where $u_{x'}$ and $u_{z'}$ are lateral and longitudinal dispacements of the 
oriented CNTs, $E'$ is the effective modulus of elasticity of CNTs, 
$A_{0}$ is the effective cross-sectional area, $A_{2}$ is the 
second moment of cross-sectional area about $Z$-axis, 
$\Delta T(z')=T(z')-T_0$ is the difference between the 
absolute temperature ($T$)
during field emission and a reference temperature ($T_0$), $\alpha$ is 
the effective coefficient of thermal expansion (longitudinal), 
$\rm C_{vs}$ is the van der Waals coefficient, superscript $(m)$ indicates 
the $m$th wall of the MWNT with $r^{(m)}$ as its radius, $\Delta_{x'}$ is the 
lateral displacement due to pressure and
$\rho$ is the mass per unit
length of CNT. We assume fixed boundary conditions ($u=0$) at 
the substrate-CNT interface ($z=0$) and forced boundary conditions at
the CNT tip ($z=h(t)$).\par
The governing equation in temperature is obtained by 
the thermodynamics of electron-phonon interaction. 
By considering the Fourier heat conduction and thermal
radiation from the surface of CNT, the energy rate balance equation 
in $T$ can be expressed as
\begin{equation}
dQ - \frac{\pi d_t^2}{4} dq_F -\pi d_t \sigma_{SB}(T^4-T_0^4) dz' = 0 \;,
\label{eq:heat2}
\end{equation} 
where $dQ$ is the heat flux due to Joule heating over a 
segment of a CNT, $q_{F}$ is the Fourier heat conduction, $d_{t}$ is the 
diameter of the CNT and 
$\sigma_{SB}$ is the Stefan-Boltzmann constant. Here, we assume the
emissivity to be unity. At the substrate-CNT interface ($z'=0$), 
the boundary condition $T=T_0$ is applied and at the tip we assign 
a reported estimate of the power dissipated by phonons exiting the CNT 
tip~\cite{bib:chiu} to the conductive flux. We first compute the electric 
field at the nodes and then solve all the governing equations 
simultaneously at each time step and the curved shape 
$s(x'+u_{x'},z'+u_{z'})$ of each of the CNTs is updated. 
The angle of orientation $\theta$ between the nodes $j+1$ and 
$j$ at the two ends of segment $\Delta s_j$ is expressed as
\begin{equation}
\theta (t) = \tan^{-1} \Bigg(\frac {(x^{j+1}+u_{x}^{j+1})
-(x^{j}+u_{x}^{j})} {(z^{j+1}+u_{z}^{j+1})-(z^{j}+u_{z}^{j})}\Bigg) \;,
\label{eq:damp51}
\end{equation}
\begin{equation}
\left[ \begin{array}{ccc}
u_{x}^{j}\\
u_{z}^{j}\\
\end{array} \right]
= [\Gamma (\theta (t-\Delta t)^j)] \left[ \begin{array}{ccc}
u_{x'}^{j}\\
u_{z'}^{j}\\
\end{array} \right] \;,
\label{eq:damp52}
\end{equation}
where $\Gamma$ is the usual coordinate transformation matrix which maps 
the displacements $(u_{x'},u_{z'})$ defined in the local $(X',Z')$ 
coordinate system into the displacements $(u_{x},u_{z})$ defined in 
the cell coordinate system $(X,Z)$. For this transformation, we employ 
the angle $\theta (t-\Delta t)$ obtained at the previous time step 
and for each node $j=1,2,3,\dots$.

\section{RESULTS AND DISCUSSIONS}

The CNT film considered in this study consists of randomly 
oriented multiwalled CNTs. The film was grown on a stainless 
steel substrate. The film surface area (projected on anode) 
is 49.93 $\rm mm^2$ and the average thickness of the film (based on randomly 
distributed CNTs) is 10-14 $\rm \mu m$. In the simulation and 
analysis, the constants B and C in Eq.~(\ref{eq:FN}) were taken as 
$B=(1.4\times 10^{-6})\times \exp((9.8929)\times \Phi^{-1/2})$ and 
$C = 6.5\times 10^{7}$, respectively~\cite{bib:huang}. 
It has been reported in the literature (e.g., ~\cite{bib:huang}) 
that the work function $\Phi$ for CNTs is smaller than the work 
functions for metal, silicon, and graphite. However, there are significant 
variations in the experimental values of $\Phi$ depending on the types of
CNTs (i.e., SWNT/MWNT), geometric parameters. The type of substrate
materials have also significant influence on the electronic
band-edge potential.  
The results reported in this paper are based on computation with
$\Phi= 2.2 eV$. \par

\begin{figure}[h!]
\begin{center}
\epsfig{file=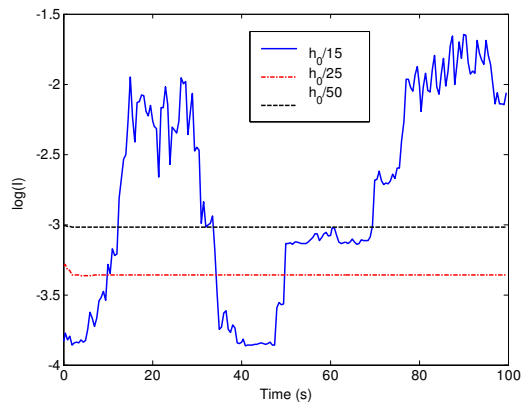, width=7cm}
\end{center}
\label{fig:It}
\caption{Field emission current histories for various initial 
average tip deflections and under bias voltage of 650V. The current 
I is in Ampere unit.}
\end{figure}

\begin{figure}[h!]
\begin{center}
\epsfig{file=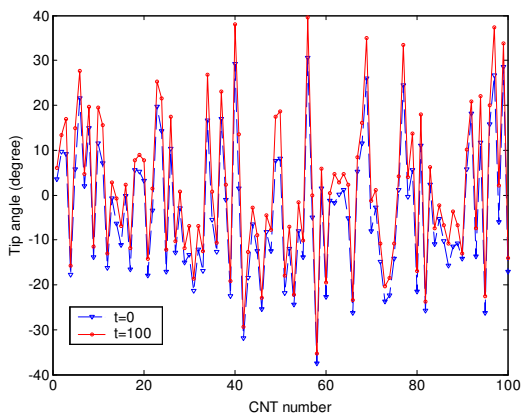, width=7cm}
\end{center}
\label{fig:angle}
\caption{Comparison of tip orientation 
angles at t=0 and t=100s.}
\end{figure}

\begin{figure}[h!]
\begin{center}
\epsfig{file=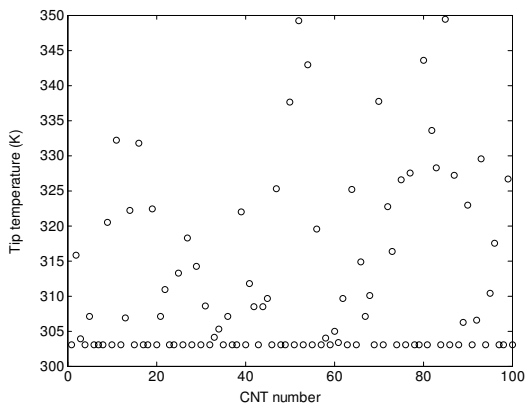, width=7cm}
\end{center}
\label{fig_temperature1}
\caption{Maximum temperature of CNT tips during 
100s of field emission.}
\end{figure}

Following sample configuration has been used in this study: 
average height of CNTs $h_{0}=12 \mu m$, uniform diameter 
$d_{t}= 3.92 nm$ and uniform spacing 
between neighboring CNTs at the substrate contact region in the film 
$d_{1}= 2 \mu m$. The initial height distribution $h$ and the orientation 
angle $\theta$ are randomly distributed. 
The electrode gap ($d$) is 
maintained at $34.7 \rm \mu m$. 
The orientation of CNTs is 
parametrized in terms of the upper bound of the CNT tip deflection 
(denoted by $h_{0}/m'$, $m'>>1$). Several computational runs 
are performed and
the output data are averaged out at each sampling time step. 
For a constant bias voltage ($650 V$ in this case), as the 
initial state of deflection of 
the CNTs increases (from $h_{0}/50$ to $h_{0}/25$), the average current 
reduces until the initial state of deflection becomes large enough 
that the electrodynamic interaction among CNTs produces sudden 
pull in the deflected tips towards the anode resulting in current 
spikes (see Fig. 3). As mentioned earlier, spikes in the 
current have also been 
observed experimentally. Fig. 4 reveals that after 
experiencing 
the electrodynamic pull and Coulombic repulsion, some CNTs reorient 
themselves. In Fig. 5, maximum tip temperature 
distribution over an 
array of 100 CNTs during field emission over 100 s duration is plotted. 
The maximum temperature rises up to approximately 350 K. \par

\section{CONCLUSION}

In this paper, a model has been developed from the device design 
point of view, which sheds light on the coupling issues related 
to the mechanics, the thermodynamics, and the process of collective 
field emission from CNTs in a thin film, rather than a single 
isolated CNT. The proposed modeling approach 
handles several complexities at the 
device scale. While the previous works by the authors mainly dealt 
with decay, kinematics and dynamics of CNTs during field emission, 
this work includes some more aspects that were assumed constant 
earlier. These include: (i) non-local nature of the electric field, 
(ii) non-linear relationship between the electronic transport and 
the electric field, and (iii) non-linear relationship between the 
electronic transport and the heat conduction. The trend in the 
simulated results matches qualitatively 
well with the results of published experimental studies.\par

\end{document}